\documentclass[conference,a4paper]{IEEEtran} 
\usepackage{indentfirst}
\usepackage{amstext}
\usepackage{amsmath}%
\usepackage{amsfonts}%
\usepackage{amssymb}%
\usepackage{graphicx}
\usepackage{epsfig}
\usepackage{graphicx}
\usepackage{enumitem}
\usepackage{mdwlist}
\usepackage{graphicx}

\usepackage{subfigure}
\usepackage{algorithm,algorithmic}
 \usepackage{array}
 \renewcommand\IEEEkeywordsname{Keywords}
   \usepackage{booktabs}

\makeatletter
\newcommand*\titleheader[1]{\gdef\@titleheader{#1}}
\AtBeginDocument{%
	\let\st@red@title\@title%
	\def\@title{%
		\bgroup\normalfont\small\centering\@titleheader\par\egroup
		\vskip0.6em\st@red@title}
}
\makeatother

\title{Enhanced Minimal Scheduling Function for IEEE 802.15.4e TSCH Networks}
\titleheader{This is the authors’ version of the paper that has been accepted for publication in IEEE Wireless Communications and Networking Conference,  15- 18 April 2019, Marrakesh, Morocco}

\begin{document}
\pagestyle{plain}
\setlength\abovecaptionskip{0.50ex}
\setlength\belowcaptionskip{0.05ex}
\author{\IEEEauthorblockN{ Taieb Hamza and Georges Kaddoum}
\IEEEauthorblockA{Department of Electrical Engineering \\
     LACIME Laboratory\\
      University of Quebec, ETS, Montreal, Canada.\\
      taieb.hamza.1@ens.etsmtl.ca, georges.kaddoum@etsmtl.ca}
}
\maketitle
\begin{abstract}
MAC layer protocol design in a WSN is crucial due to the limitations on processing capacities and power of wireless sensors. The latest version of the IEEE 802.15.4, referenced to as IEEE 802.15.4e,  was released by IEEE and outlines the mechanism of the Time Slotted Channel Hopping (TSCH). Hence, 6TiSCH working group has released a distributed algorithm for neighbour nodes to agree on a communication pattern driven by a minimal scheduling function. A slotframe contains a specific number of time slots, which are scheduled based on the application requirements and the routing topology. Sensors nodes use the schedule to determine when to transmit or to receive data. However, IEEE 802.15.4e TSCH does not address the specifics on planning time slot scheduling.
In this paper, we propose a distributed Enhanced Minimal Scheduling Function (EMSF) based on the minimal scheduling function, which is compliant with 802.15.4e TSCH. In this vein, we introduce a distributed algorithm based on a Poisson process to predict the following schedule requirements. Consequently, the negotiation operations between pairs of nodes to agree about the schedule will be reduced. As a result, EMSF decreases the exchanged overhead, the end-to-end latency and the packet queue length significantly. Preliminary simulation results have confirmed that EMSF outperforms the 802.15.4e TSCH MSF scheduling algorithm. 
\end{abstract}
\IEEEpeerreviewmaketitle
\section{Introduction}
The next industrial revolution is announced to be Industry 4.0, which will reduce cost and maximize flexibility with the use of digital automation \cite{1}. The Industrial Internet of Things (IIoT) plans to connect to the Internet a large number of industrial objects. In this regard, it is mandatory for real-time infrastructure to be highly reliable for wireless transmissions. So far, the efforts undertaken in IoT were for best-effort solutions, but industrial applications demand a higher level of control in terms of reliability and delay \cite{2}. For that reason, specific MAC protocols are needed to add strict guarantees. Deterministic approaches, especially, can be used to allocate a fixed bandwidth to every device or flow. In addition, these approaches can isolate flows, where each type of flow gets a specific transmission bandwidth. In IEEE 802.15.4e TSCH mode, channel hopping mechanism is used to reduce external interference and fading. TSCH deploys a deterministic approach to avoid collisions by careful scheduling, which involves allocating a group of cells for interfering transmitters to avoid collisions while reducing contention. A number of centralized and distributed scheduling algorithms have been introduced as of date for TSCH \cite{3}. Over-provisioning is a method used to cope with unreliable links in the schedule by reserving a few cells to allow a packet to be retransmitted. However, extra cells leads to an increase in delays and jitter \cite{4}. Network capacity is also reduced if there is too much traffic and/or lack of reliability due to external interference. In order to handle the issues concerning scheduling, routing and internet integration, the IETF IPv6 through the TSCH mode of IEEE802.15.4e was established for an improved integration of the IPv6-enabled protocols such as RPL, 6LoWPAN and CoAP \cite{15}. In this paper, we propose an Enhanced Minimal Scheduling Function (EMSF) in order to improve reliability and latency when pairs of nodes are dynamically determining their bandwidth requirements. The proposed algorithm is based on the minimal scheduling function (MSF) schemed for the 6TiSCH networks. Our proposal is designed to meet the following goals: 
\begin{enumerate}
\item A dynamic scheduling: The proposed scheme allocates and deallocates cells without considering the threshold-based mechanism.
\item A reduction of the transmitted data during the scheduling negotiation phase: The EMSF is designed to meet this target by introducing a prediction system model which anticipates the data to be transmitted for each pair of nodes in the next slotframe. 
\item A reduced latency: This design is introduced by minimizing the overhead control packets in order to diminish the end-to-end data transmission delay and the queued data.
\end{enumerate}
The remainder of this paper is organized as follows. In section II, we provide an outlook to the 6TiSCH WG over the 802.15.4e TSCH mode. We discuss some background information
and related work in section III. The design of the system is described in section IV and V. We illustrate, in section VI, the results of simulations that show the performance of our approach, and we conclude the paper in section VII.
\section{IEEE 802.15.4e TSCH}
\subsection{Concept of IEEE 802.15.4e TSCH}
The nodes in IEEE 802.15.4e TSCH network communicate using a time-slotted mechanism over multiple frequencies, which follows a Time Division Multiple Access (TDMA) schedule. It partitions the wireless spectrum into time and frequency, which is scheduled over a set period of time. This scheduling is also called a superframe or a slotframe. A node transmits/receives data to/from its neighbours on a predefined timeslot and channel in the schedule. A cell, which is usually 10ms in length, is a basic unit of bandwidth scheduled. The transmitter in a cell sends a data packet to the receiver. Once successfully received, an acknowledgement is sent back by the receiver. Channel-hopping improves communication by making it more reliable through diversification of frequencies. This statistical mitigation reduces narrow-band interference and multi-path fading. Cells can contain multiple communication links as long as they are not conflict links nor links interfering amongst each other. Conflict links are those that have the same receiver and/or transmitter. The communication links bounce over a series of available channels in a quasi-random way between the super-frames. Both the sender and the receiver for each scheduled cell will use Eq.(\ref{eqq1}) to calculate communication frequency, i.e., \textit{f} \cite{for1}:
\begin{equation}
    f=\rm{\textit{F}\{(\textit{ch}_{offset}+\rm{ASN}) \,\rm{mod}  \, \textit{N}_{ch}\}}.
    \label{eqq1}
\end{equation}
where $F$ is the mapping function for channel frequency, $ch_{\text{offset}}$ is the channel offset, ASN is the total number of timeslots, and finally mod $\, N_{ch}$ is the modular division of $N_{ch}$, which refers to the number of available physical channels.
ASN is calculated as follows:
\begin{equation}
    \rm{ASN=}\textit{K} \times \textit{S}+ \textit{T}.
\end{equation}
where $K$ is the slotframe cycle, $S$ is the size of the slotframe, and $T$ are the allocated timeslots.
\subsection{Scheduling in 6TiSCH networks}
Recently, 6TiSCH Wireless Group was working to define a pre-configured or learned minimal schedule of a node that joins a network with static scheduling configuration \cite{5}. Various kinds of frame
are transmitted and received through cells within a schedule determined by slotted ALOHA protocol. Note that static scheduling within 6TiSCH is used only for specific situations like
during the bootstrapping stage or as a fallback during network failures. For other situations, a scheduling function is used in order to allocate or deallocate cells between neighboring
nodes. 6TiSCH uses MSF as a default scheduling function presented in the IETF draft \cite{6}. The allocation policy and the bandwidth estimation algorithm are used by MSF to determine when cells in neighbouring nodes should be added or deleted \cite{7}. In addition, MSF uses a threshold-based mechanism to mitigate against sudden fluctuations and increases in bandwidth by adding or deleting operations. Both add and delete negotiations are similar, however. Despite that, given two nodes A and B in a network, node B contains all slotOffsets of node A candidate cells of the remove CellList requests. Since candidate cells are randomly selected, there is a high likelihood of negotiation errors at node A. This is due to the fact that node B contains fewer cells in CellList with available slotOffsets than NumCells. In addition, other pairs of nodes may use allocated cells that are also used by nodes A and B. This may cause network collisions, which are mitigated by the scheduling function. The minimal scheduling function uses 6top Housekeeping function to track cell usage and performance and to relocate cells that have collided \cite{8}.
\section{Related works}
Constructing a schedule is application specific and several scheduling algorithms have been introduced to schedule TSCH networks. Different proposals could be used to set up the schedule. They can be classified as centralized and distributed. In centralized approaches, the DAG root builds and maintains the schedule for the entire network. Palattella et al. proposed Traffic Aware Scheduling Algorithm (TASA), a centralized scheduling technique that uses a leading-edge matching and colouring method of graph theory to map the distribution of time slots and channel offsets \cite{9}. The motes in the network send the requirements of bandwidth and  energy, TASA computes a schedule that satisfies those requests and returns them back to the motes. In distributed approaches, the nodes negotiate with their neighbors to build their own schedule.  Accentura et al. proposed Decentralized Traffic Aware Scheduling (DeTAS), which builds optimal collision-free multi-hop schedules \cite{10}. DeTAS employs neighbour-to-neighbour signaling in order to gather network and traffic information. It ensures the smallest queue utilization and the shortest possible end-to-end latency period between when the data was generated to when it was received. Orchestra is another non-graph scheduling technique in which nodes compute their own local schedules, hence the reason it’s referred to as autonomous scheduling of the TSCH in IPv6 Routing Protocol for Low-Power and Lossy Networks (RPL) \cite{11}. It has no central entity nor negotiation and allocates slots such that it can be installed or deleted automatically with the evolution of the RPL topology. Despite that, it does not offer any solution against bursty traffic. Domingo-Prieto et al. introduced a distributed scheduling algorithm as a solution against sudden or bursty traffic, which uses a control paradigm called Proportional Integral and Derivative (PID) \cite{12}. Techniques based on graph theory have been introduced to other networks including peer-to-peer networks and cognitive radio networks. It is worth mentioning that the work of Domingo-Prieto et al. is the first one to use graph-theory based combinatorial properties to address scheduling in IEEE 802.15.4e TSCH networks. Soua et al. proposed the Wave scheduling algorithm \cite{13}. It aims to reduce the slotframe size by dividing it with a unit called wave and achieving waves many times to build a schedule for all the nodes at proper intervals of time to the DAG root. Every node with a transmittable packet is allocated both a timeslot and a channel for every wave. The slot or channel pattern is determined by the first wave. The following waves are modeled after the first wave, however only the transmission-containing slots and the order in which they repeat themselves are copied from the first wave. Determining a schedule is application-specific and it's based on the metrics that need to be improved. The proposed 6TiSCH WG minimal scheduling function \cite{5} defines how nodes add or remove cells based on a bandwidth estimation algorithm. This method is based on exchanging control packets to measure the required bandwidth. Hence, more resources are appealed to determine the appropriate measurements which leads to an increased probability of dropped packets and a high latency. Note that, unlike the existing scheduling approaches, our method will guarantee a minimum negotiation overhead which will reduce the latency, the packet queue size and the resource utilization. 
Our proposed model is based on three indispensable steps:
\begin{enumerate}

\item Computation of the average generated packets in the previous slotframes
\item Prediction of the amount of data (event/periodic).
\item 6p add/remove transaction according to the predicted model.
\end{enumerate}

\section{Proposed system }

In this section, we will introduce some definitions and assumptions of the system model. When a 802.15.4e TSCH node detects a substantial unpredictable physical event, a massive flow of data packets is generated and queued by the node. These nodes check if they have sufficient bandwidth to send these packets to their parents. A limited number of transmissions and re-transmissions are allowed by each node, which, when exceeded, cause packet dropping. This appliance triggers a tremendous number of packet transactions and a spike in resources usage. To reduce negotiation errors, number of dropped packets and end-to-end latency times, we propose a novel scheduling function based on the minimal scheduling function presented in the IETF draft \cite{6}.  The proposed mechanism consists of the following two main operations: Computation of the mean of packets generated by each node and the prediction of the required cells in the next slotframe.
First, we introduce some definitions:
\begin{itemize}
    \item \textbf{Definition 1:} We focus on event-driven WSN where nodes measure physical events and send it to the sink in an upstream data circulation transfer.
    \item \textbf{Definition 2:} We define the network topology as graph $G = (V,E)$ where $V$ is the set of all nodes and $E $ is the set of edges between the nodes displaying symmetric communications links.
    \item  \textbf{Definition 3:} In a data gathering frame of any node $n$ we denote \rm{G}(\textit{n}), which is the number of data transmitting packets sent by node \textit{n} and T(\textit{n}), which is the sum of all the packets sent by \textit{n} including $G(n)$ and the number of packets received by the parent from its children. Therefore, we define \textit{T(n)} using the following equation:
\begin{equation}
   T(n)=\sum_{v \rm{\in subTree(n)}}G(V).
\end{equation}
\item \textbf{Definition 4:} We define $\rm{Q}_C^P(n)$ as the number of packets in the queue of $n$ that has to be sent to parent and $\rm{C}_C^P(n)$ the number of cells already allocated between the parent $\rm{p}(n)$ and the child $\rm{c}(n)$.
\item  \textbf{Definition 5:} After executing the scheduling algorithm and based on its output, nodes in $G $ will either add, delete, or keep cells in the next slotframe $S_{i+1}$.
\end{itemize}
In this paper we adopt the following assumptions:
\begin{itemize}
    \item \textbf{Assumption 1:} We assume that the gathered data network topology and the routing tree are provided.
    \item \textbf{Assumption 2:} We consider also that the links of the routing tree are symmetric, since user data is gathered upstream, whereas the schedules are negotiated in a distributed fashion between nodes.
    \item \textbf{Assumption 3:} Symmetric links are a requirement for the instant acknowledgment policy.
\end{itemize}
\section{Enhanced minimal scheduling algorithm}

\subsection{Prediction of the amount of data }
In 802.15.4e TSCH networks, a node transmits two types of packets: Periodic data $A_i^w(n)$ and Event-driven data $A_i^v(n)$. Periodic data consist of enhanced beacons, which contain information about the actual ASN, the length of the timeslot and other information about the network. Event-driven data are sent upon the detection of a physical event. The total of a throughput a node generating is illustrated by the following formulas:
\begin{equation}
    A_i^T(n)= A_i^w(n)+ A_i^v(n).
\end{equation}
Which is equivalent to:
\begin{equation}
    A_i^T(n)= \sum_{i-1}^i \rm{subTree}(n)+ \sum_{i-1}^i \textit{G(n)},   \rm{i<1<sink}.
\end{equation}
\subsection{ Poisson-based packet generation model}
As mentioned in the last section, we considered the 802.15.4e TSCH network as an event-driven network, where sensor nodes report data to the sink only when they obtain new data (an event occurs in a sensing area). We formulate the scheduling problem as a Poisson process model to describe the data generation.\\
\textbf{Process Poisson Definition  \cite{poisson}:}
A Poisson process ${N(t),t \ge 0}$ with intensity $\lambda > 0$ is a counting process with the following properties.
\begin{enumerate}
    \item \textbf{ Independent increment}: For all $t_0 = 0 < t_1 < t_2 < · · · <t_n$, then $N(t_1) - N(t_0), N(t_2) - N(t_1), . . . , N(t_n) - N(t_{n-1})$ are independent random variables.
\item \textbf{ Stationary increments with Poisson Distribution }: For all
$s \ge 0,t > 0, N(s + t) - N(s) \sim Poisson(\lambda t)$.
\end{enumerate}

The system satisfies the previous Poisson lemmas explained as follows:\\
    Lemma 1: Events are considered as independent (temporally 
    \begin{figure*}[!ht]
	\centering
	\subfigure[Average generated packets] {\includegraphics [height=5.cm, width =6.3cm]{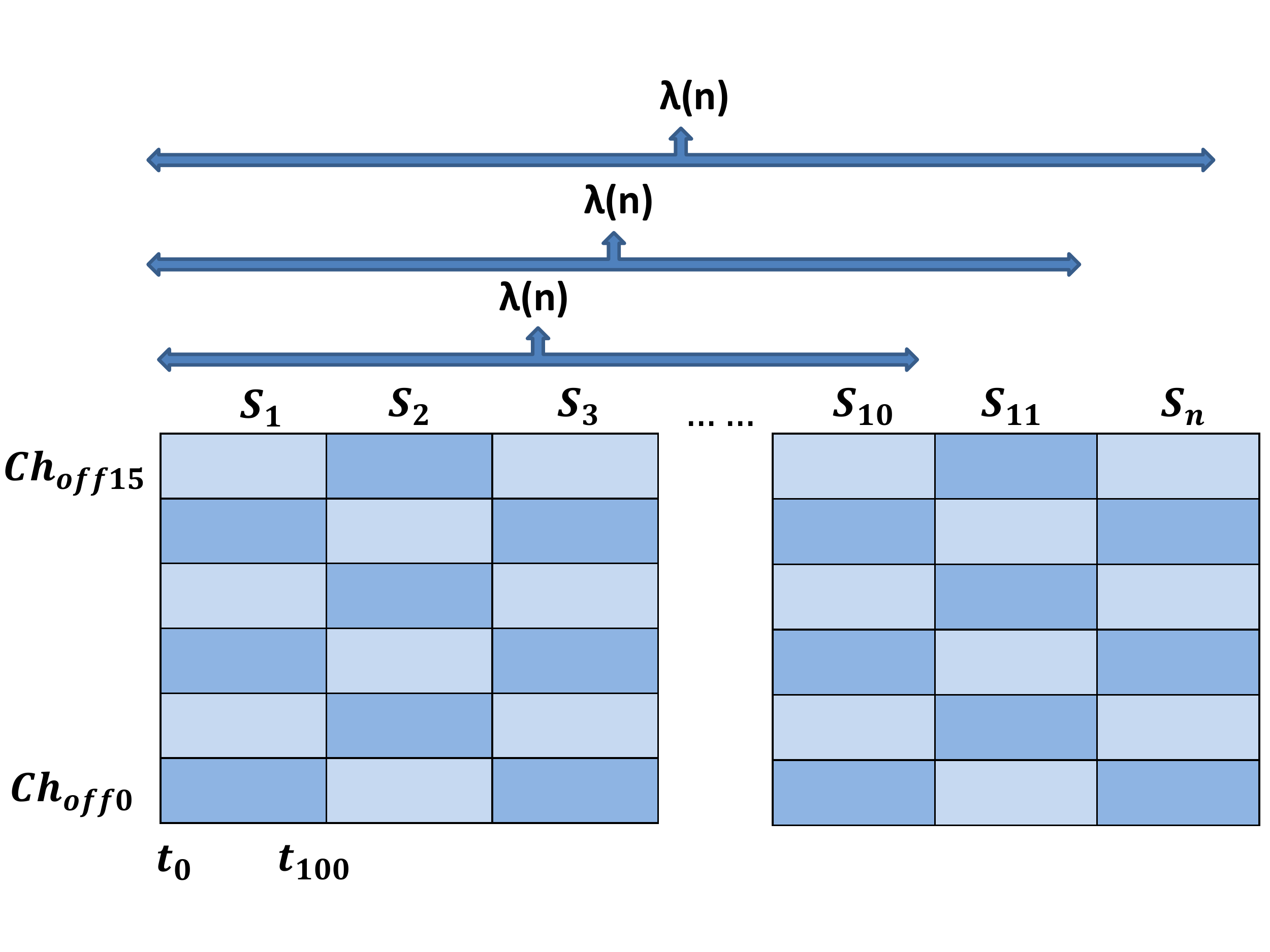}}
	\label{aa}
	\subfigure[Poisson prediction model]{\includegraphics [height=5.cm, width =6.3cm]{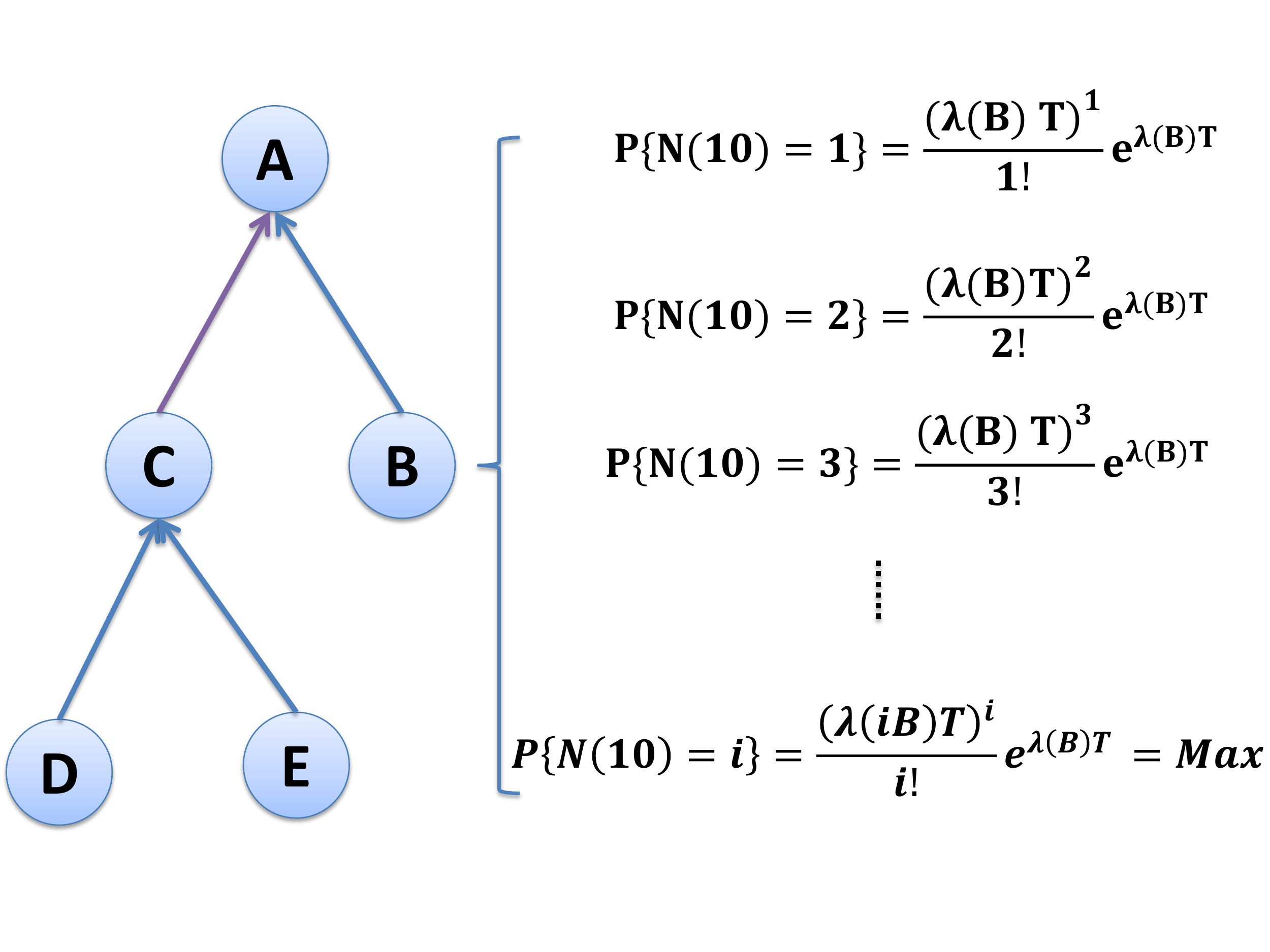}}
	\subfigure[6p operations]{\includegraphics [height=5.cm, width =5cm]{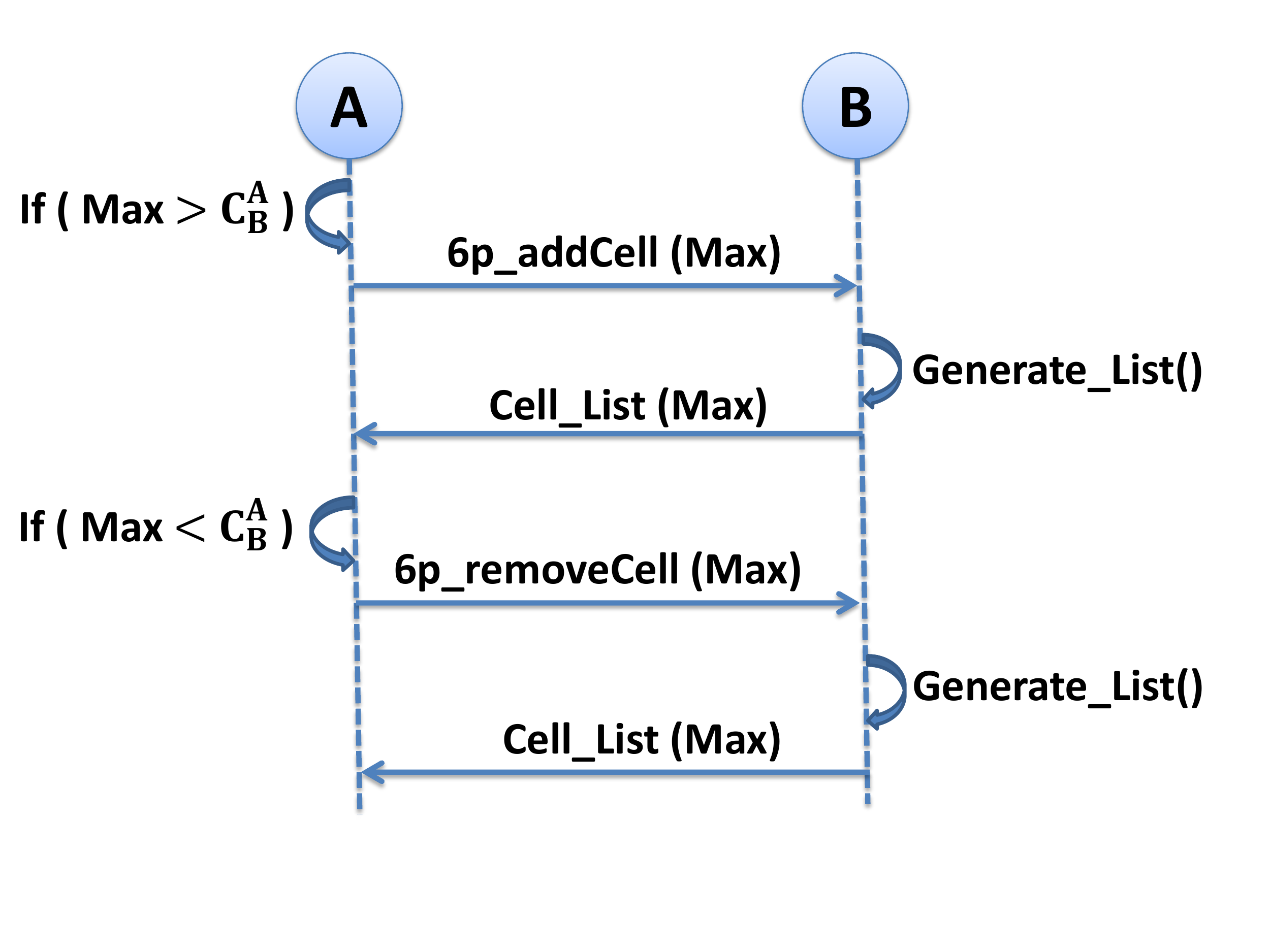}}
	\caption {System model}
	\label{system}
	\end{figure*}
    and spatially) and occur with equal probability over the area. In the network, we consider the generated packets by different nodes as an independent event with no relation with the preceding event.\\
    Lemma 2: The Time  \textit{T} between each slotframe \textit{S} is finite (slotframe durations stay the same through all the scheduling).
Based on  lemma (1) and (2), we can adopt a Poisson process model to describe data generation packets in the network. The distribution of the number of data packets, $N(T)$, generated by each node in the network from the beginning to the end of the slotframe is equal to the following formula: 
\begin{equation}
\label{form1}
    P\{N(t)=n\}= \frac{(\lambda t)^n}{n!} e^{\lambda t}.
\end{equation}

where $\lambda$ is the average number of packets generated after a certain number of slotframes have passed since the two nodes are synchronized to the schedule. Mathematical, $\lambda$ is defined by the following formula:
\begin{equation}
\label{form2}
    \lambda(n)=\frac{\sum_{i=T-\beta}^T \rm{nbPacket_i(n)}}{\beta}.
\end{equation}

where \textit{T} is the actual instant, \rm{$nbPacket_i(n)$} is the number of packets generated by the node $n$ in the instant $T=i$ and $\beta$ is the sum of previous number of slotframes which can be validated through simulations.
In order to get an accurate value of $\lambda $, the protocol will execute it's scheduling algorithm with the minimal scheduling function up to $\beta=10$ slotframes.

An illustration of the execution of the system model is described in Fig. \ref{system}. As an example, the nodes A and B are exchanging packets over the built scheduling by the minimal scheduling function. Starting from slotframe \textit{S} = 10, using Eq.(\ref{form2}), node B determines the average number of generated packets in the previous slotframes since it joined the network as it is shown in Fig. 1(a). Then, the average number is used in Eq.(\ref{form1}) in order to determine the probability of getting a determined number of packets  as depicted in Fig. 1(b). This operation will be repeated until a maximum probability value is reached. Based on the predicted output as observed in Fig. 1(c), the algorithm will execute either an add or a remove 6p operation explained in the following section.

\subsection{Add/remove cells}
Starting from a given slotframe cycle, each node in the network will execute the Algorithm \ref{algo}. 
The purpose is to predict the number of packets that will be generated by each node in the next slotframe. To reduce resource computation by each node, the algorithm stops when a maximum value of $\lambda $ is reached. By knowing the number of packets that will be generated in the next slotframe, a node can predict how many cells are required in order to exchange data with its preferred parent. Based on the output of the algorithm, a node can trigger a 6p transaction with its preferred parent either to add, or remove cells to the TSCH schedule of both nodes. 

\begin{algorithm}[!h]
\caption{Cell prediction algorithm {Starting from Slotrame S=$\beta$}}
\begin{algorithmic}[1]
\STATE Set \textit{G(n)} to be a random value chosen from a Poisson distribution with mean $\lambda =rt$ (r in unit of 1/time)
 \\
\FOR {$S= S_{11}$ to \{Network lifetime\} ; $S_{i++}$}
 \STATE Determine $\lambda$ using Eq.(\ref{form2})  $     \triangleright$\{Average generated number of packets in $S_0$, $S_1$, $S_2$,..., $S_{i-1}$ \}
 \STATE	$\lambda \Longleftarrow$ $lambda(n)$

\WHILE{$p<max$ } 
\STATE $p \Longleftarrow$ Determine  $(PN(t), \lambda)$ $     \triangleright$\{probability of generating $\lambda$ packets\}
\IF {($p \geqslant pmax$)}
\STATE $max \Longleftarrow p$ 
\ENDIF
\ENDWHILE
\STATE return (p) $     \triangleright$ \{the maximum value of the probability\}
\STATE \textbf{if} ($p < C_c^p(n)$) $     \triangleright$\{Compare p with the actual reserved cells for node n\} \textbf{then}
\STATE  6P\_DELETE\_command (p) 	$     \triangleright$\{6P delete request of $C_c^p(n)$ - p slots\}
\STATE \textbf{elseif} ($p > C_c^p(n)$) \textbf{then}
 \STATE	6P\_ADD\_command (p) 	$     \triangleright$\{6P add request of $C_c^p(n)$ + p slots\}
\STATE \textbf{end if} 
\ENDFOR
\end{algorithmic}
\label{algo}
\end{algorithm}

\section{Performance Evaluation }
   
We conducted our simulations on OpenWSN, which is an opensource network simulator for wireless sensor networks. The simulator supports a few major types of protocol based on IoT standards, such as 6LoWPAN, RPL, ROLL, and CoAP. All the protocols based on OpenWSN 6TiSCH use the latest IEEE 802.15.4e TSCH standard in order to improve their reliability and stability \cite{14}.
In order for OpenWSN to remain up to date and have the desired network metrics, it provides a python-based configuration tool that allows its network parameters to be modified. The behavior of the network (such as PDR in every channel), the node queuing priority and the timestamp of each sending and receiving packet is recorded and then simulated. Table \ref{tabb} shows the parameters used in the OpenWSN simulation.
\begin{table}[!h]
\caption{OpenWSN Simulation parameters}
\label{tabb}
\center
\begin{tabular}[19.cm]{|c|c|}
\hline
\textbf{Simulation Parameters}&\textbf{Values}\\
\hline
Available channels ($N$) &11-26\\
\hline
Number of nodes &2-100\\
\hline
Timeslot duration&10 ms\\
\hline
Slotframe length $p$ &101 timeslot\\
\hline
Payload &127 bytes\\
\hline
MAC max retries& 4\\
\hline
Max queue size& 5\\
\hline
Period of transmission& 200 ms\\
\hline
\end{tabular}
\end{table} 
\subsection{6p error ratio}

\begin{figure}[!h]
  \centering
    \includegraphics[width=7.5cm]{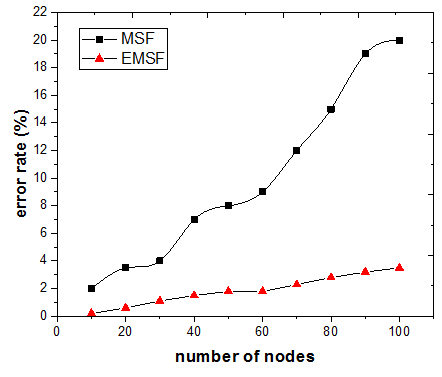}
  \caption{6p error ratio}
  \label{error}
\end{figure}

6p negotiation error ratio is the average ratio between the number of 6p transaction errors and the total 6p transactions throughout a slotframe cycle. Fig. \ref{error} displays the way that the 6p negotiation error ratio is affected by the network density. The number of nodes passes from 10 to 100. 
The negotiation error ratio increases with the increase in network density. In the MSF scenario, it goes from  2.1\% to 19.8\%, whereas in EMSF scenario it goes from 0.2\% to 3.5\%. This is due to the decrease in the number of control packets and to the substitution of the threshold-measurement mechanism that is needed for the prediction system to add or delete  6p operations. The number of 6p exchanged packets decreases, which leads to a lower negotiation-error ratio. The suggested mechanism, the EMSF, largely surpasses the MSF and keeps the negotiation-error ratio under 3.5\% for all of the network densities.

\subsection{Packet overhead}

\begin{figure}[!h]
  \centering
    \includegraphics[width=7.8cm]{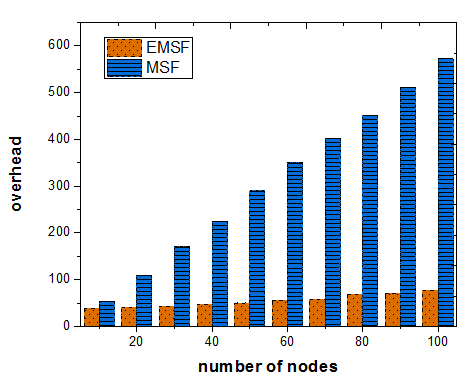}
  \caption{packet overhead}
  \label{over}
\end{figure}

Fig. \ref{over} represents the overhead traffic load in bytes that was employed by nodes in order to exchange 6p information in the network. We notice that the number of exchanged messages increases linearly with the number of deployed nodes. This is due to negotiation exchanges taken between pairs of nodes to determine the 6p operation that will be deployed.
We notice that the EMSF maintains an almost constant amount of exchanged packets. This is due to the prediction algorithm that anticipates the number of required cells for each pair of nodes in the following slotframe. EMSF avoids sending overloads and keeps a constant average of control packets through the network lifetime.

\subsection{Latency}

\begin{figure}[!h]
  \centering
    \includegraphics[width=7.9cm]{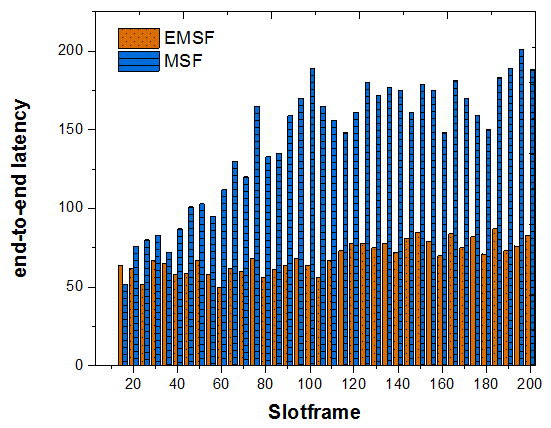}
  \caption{Latency}
  \label{latency}
\end{figure}

The latencies are calculated as follows: Every packet is timestamped since it is generated in the application layer of a source node, until it reaches the application layer of the DAG root. As a result of this condition, retransmission on a MAC layer is not taken into consideration. However, if a packet is repeatedly transmitted, a peak in latency may occur. When a packet achieves 4 MAC retransmissions attempts and the queue is full, a packet will be considerably dropped. Each node sends a fixed traffic load of 2 packets per slotframe for the first 50 slotframes. Thereafter, each node sends a sporadic traffic load ranging from 2 to 7 packets per slotframe. 20 nodes were deployed in this simulation. An end-to-end latency comparison between MSF and EMSF is  illustrated in Fig.\ref{latency}. The latency was almost constant for the first 50 slotframes because of stable transmitted data flow. Thereafter, latency varies in between cycles as a result of the failing stochastic transmissions based on the PDR between the nodes. EMSF keeps an end-to-end latency below 75 milliseconds in almost all the slotframe cycles. This is due to its scheduling mechanism that minimizes the overhead charges, which increases the PDR even when irregular data flow is generated. 
\subsection{ Node queue size}

The following Fig. \ref{queue} represents the average number of packets in the queues of the network nodes. From this figure, it can be seen that in the case of MSF, node queues are almost full and reach their maximum (5 packets) in some slotframes causing more dropped packets. On the other hand, EMSF shows better performance in this concern. 
\begin{figure}[!h]
\centering       
 	\subfigure[EMSF]{\includegraphics[width=4.3cm]{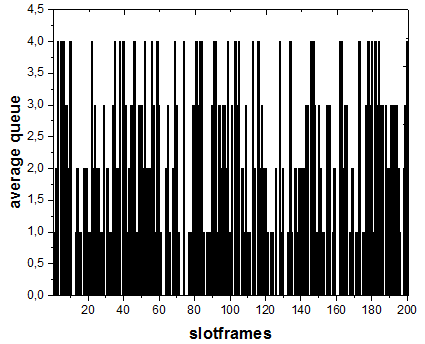}\label{fig:random}}
    \subfigure[MSF]{\includegraphics[width=4.3cm]{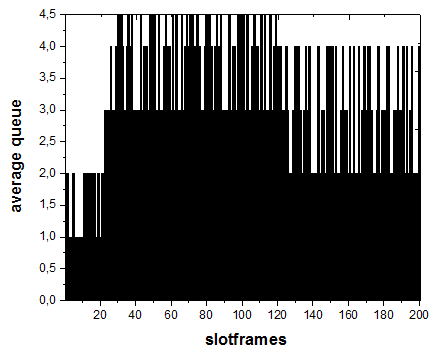}\label{fig:circular}}   
	\caption{Node queue size}
	\label{queue}
\end{figure}
Throughout 200 slotframes, EMSF shows an average queue size ranging from 0 to 4 packets. In a lot of cases, nodes over-allocate cells, especially those near to the DAG root. This is due to the prediction algorithm in the case where a high packet rate is generated in the previous slotframes. This will allocate more cells between pairs of nodes, which will be used when an unpredictable event occurs. For instance, given an average number of generated packets equal to 10 in the previous slotframes, a node generating 5 packets will allocate cells based on the average number of packets generated since the time it joined the network. This means that, even though a huge number of packets is generated, the node has already reserved more cells than its requirements. Therefore, packets tend to queue up less, but consume more energy and than in the case of MSF.

\section{Conclusion}

In this paper, we formulated the scheduling function for IEEE 802.15.4e TSCH networks as a Poisson process model. We demonstrated that, based on the prediction algorithm, scheduling negotiation overhead is decreased noticeably. In addition, simulations approved that EMSF outperforms MSF in terms of the average packet queue length in the network and the end-to-end latency. In a future work, we will improve EMSF by combining it with an enhanced cell selection mechanism in order to reduce energy consumption throughout the network.


\begin{thebibliography}{1}


\bibitem{1}X. Li, D. Li, J. Wan, V. Vasilakos, C. Lai, S. Wang, \emph{ A review of industrial
wireless networks in the context of Industry 4.0}, \textit{Wireless Netw.}, vol. 23, no. 1, pp. 23-41, Nov. 2017.
\bibitem{2} I. Al-Anbagi, M. Erol-Kantarci, T. Mouftah, \emph{ A survey on cross-layer quality of-service approaches in WSNs for delay and reliability-aware applications}, \textit{IEEE Commun. Surveys Tuts. }, vol. 18, no. 1, pp. 525-552, Oct. 2016 .


\bibitem{3} R. Hermeto, A. Gallais, F. Theoleyre, \emph{ Scheduling for IEEE802.15.4-TSCH and slow channel hopping MAC in low power industrial wireless networks: a survey}, \textit{Comput. Commu.}, vol. 114, pp. 84-105, Dec. 2017.


\bibitem{4} Georgios Z. Papadopoulos,  T. Matsui,  P. Thubert,  G. Texier, T. Watteyne,  N. Montavont,
 \emph{ Leapfrog collaboration: toward determinism and predictability in industrial-
IoT applications}, \textit{ in Proc. Int. Conf. Commun. (ICC 2017)}, Paris, pp. 1–6, Jul. 2017.
\bibitem{15} 6TiSCH working group, https://datatracker.ietf.org/wg/6tisch/charter/, (Accessed 18 Sept. 2018).
\bibitem{for1} 6TiSCH working group, https://tools.ietf.org/html/rfc7554 (Accessed 20 Sept. 2018).

\bibitem{5} X. Vilajosana, K. Pister, \emph{ Leapfrog collaboration: toward determinism and predictability in industrial-
IoT applications}, \textit{in Proc. Int. Conf. Commun. (ICC 2017)}, Paris, Jul. 2017. 

\bibitem{6} D. Dujovne, \emph{6tisch 6top Scheduling Function Zero  (SF0) (work in progress)}, \textit{Ietf draft}, 2017.

\bibitem{7} 
M. Palattella,  T. Watteyne, Q. Wang, K. Muraoka, N. Accettura, D. Dujovne,  L. Grieco, T. Engel, \emph{ On-the-fly bandwidth reservation for 6TiSCH wireless industrial networks}, \textit{IEEE Sensors J.}, vol. 16, no. 2, pp. 550-560, Jan. 2016 


\bibitem{8}M. D. Prieto, T. Chang,  X. Vilajosana, T. Watteyne, \emph{Distributed pid-based scheduling for 6tisch networks}, \textit{IEEE Commun. Lett.}, vol. 20, no. 5, pp. 1006-1009, Mar. 2016.

\bibitem{9} M. R. Palattella, N. Accettura, M. Dohler, L. A. Grieco, and G. Boggia, \emph{Traffic aware scheduling algorithm for reliable low-power multi-hop ieee 802.15.4e networks}, \textit{in Proc. Personal Indoor and Mobile Radio Communications (PIMRC)}, Australia, pp. 327-332, Sept. 2012.

\bibitem{10} N. Accettura, M. R. Palattella, G. Boggia, L. A. Grieco, and M. Dohler, \emph{Decentralized Traffic Aware Scheduling for multi-hop Low power Lossy Networks in the Internet of Things}, \textit{ in Proc. Int. Symp on World of Wireless, Mobile and Multimedia Netw.  (WoWMoM 2013)}, Spain, pp. 1–6, Jun. 2013.

\bibitem{11} S. Duquennoy, B. Al Nahas, O. Landsiedel, and T. Watteyne, \emph{Orchestra: Robust mesh networks through autonomously scheduled TSCH}, \textit{in Proc. Conference on Embedded Networked Sensor Syst.}, Seoul, pp. 337–350, Nov. 2015.

\bibitem{12} M. Domingo-Prieto, T. Chang, X. Vilajosana, and T. Watteyne, \emph{Distributed
pid-based scheduling for 6tisch networks}, \textit{IEEE Commun. Lett.}, vol. 20, no. 5, pp. 1006-1009, Mar. 2016.

\bibitem{13} R. Soua, P. Minet, E. Livolant,\emph{ A distributed joint channel and slot assignment for convergecast in
wireless sensor networks}, \textit{ in Proc. Int. Conference on New Technologies, Mobility and Security (NTMS)}, Dubai, May 2014.
\bibitem{poisson} W. Paul, J. Baschnagel,\emph{Stochastic processes}, springer, 2013.
\bibitem{14}T. Watteyne,  X. Vilajosana, B. Kerkez, F. Chraim, K. Weekly, Q. Wang, S. Glaser, K. Pister, \emph{ Openwsn: a standards-based lowpower wireless development
environment}, \textit{ IEEE Trans. Emerging Telecommun. Technol.},  
vol. 23, no. 5, pp. 480-493, Aug. 2012.






\end{thebibliography}
\end{document}